\documentclass[aps,showpacs,preprintnumbers,amsmath, amssymb]{revtex4}

\usepackage{float}
\usepackage{graphics,epsfig}
\usepackage{graphicx}
\usepackage{dcolumn}
\usepackage{bm}

\begin{document}
\baselineskip=0.8 cm
\title{{\bf Quasinormal modes of a black hole in the deformed
Ho\v{r}ava-Lifshitz gravity}}

\author{Songbai Chen}
\email{csb3752@163.com} \affiliation{Institute of Physics and
Department of Physics,
Hunan Normal University,  Changsha, Hunan 410081, P. R. China \\
Key Laboratory of Low Dimensional Quantum Structures and Quantum
Control (Hunan Normal University), Ministry of Education, P. R.
China.}

\author{Jiliang Jing }
\email{jljing@hunnu.edu.cn}
 \affiliation{Institute of Physics and
Department of Physics,
Hunan Normal University,  Changsha, Hunan 410081, P. R. China \\
Key Laboratory of Low Dimensional Quantum Structures and Quantum
Control (Hunan Normal University), Ministry of Education, P. R.
China.}

\vspace*{0.2cm}
\begin{abstract}
\baselineskip=0.6 cm
\begin{center}
{\bf Abstract}
\end{center}

We study the quasinormal modes of the massless scalar perturbation
in the background of a deformed black hole in the
Ho\v{r}ava-Lifshitz gravity with coupling constant $\lambda=1$. Our
results show that the quasinormal frequencies depend on the
parameter in the Ho\v{r}ava-Lifshitz gravity and the behavior of the
quasinormal modes is different from those in the
Reissner-Norstr\"{om} and Einstein-Born-Infeld black hole
spacetimes. The absolute value of imaginary parts is smaller and the
scalar perturbations decay more slowly in the deformed
Ho\v{r}ava-Lifshitz black hole spacetime. These information can help
us understand more about the Ho\v{r}ava-Lifshitz gravity.

\end{abstract}

\maketitle
\newpage
\vspace*{0.2cm}

Inspiring by the Lifshitz model, Ho\v{r}ava \cite{ho1} proposes
recently a field theory model for a UV complete theory of gravity,
which is a non-Lorentz invariant theory of gravity in $3+1$
dimensions. Unlike Einstein gravity, it is renormalizable by
power-counting arguments. Thus, it is believed widely that it could
be a candidate for Einstein's general relativity. Very recently, the
Ho\v{r}ava-Lifshitz gravity theory has been intensively investigated
in \cite{ho2,ho3,VW,klu,Nik,Nas,Iza,Vol,CH,CHZ} and its cosmological
applications have been  studied in \cite{cal,TS,muk,Bra,pia,gao,KK}.
Some static spherically symmetric black hole solutions have been
found in Ho\v{r}ava-Lifshitz theory \cite{CY,KS,LMP,CCO,CLS,Gho} and
the associated thermodynamic properties with those black hole
solutions have been investigated in \cite{MK,Nis,CCO1,Myung}.

The four-dimensional metric in the ADM formalism can be expressed as
\cite{adm}
\begin{eqnarray}
 ds_{ADM}^2= - N^2 dt^2 + g_{ij} \Bigg(dx^i - N^i dt\Bigg)
\Bigg(dx^j - N^j dt\Bigg)
\end{eqnarray}
and the Einstein-Hilbert action is given by
\begin{eqnarray}
\label{Eins} S_{EH} = \frac{1}{16\pi G} \int d^4x \sqrt{g} N
\Bigg(K_{ij} K^{ij} - K^2 + R - 2\Lambda\Bigg),
\end{eqnarray}
where $G$ is Newton's constant and $K_{ij}$ is extrinsic curvature
which takes the form
\begin{eqnarray}
K_{ij} = \frac{1}{2N} \Bigg(\partial_t g_{ij} - \nabla_i N_j -
\nabla_j N_i\Bigg).
\end{eqnarray}
In general, the IR vacuum of this theory is anti de Sitter (AdS)
spacetimes. To obtain a Minkowski vacuum in the IR sector, one can
modify the theory by introducing ``$\mu^4R$" and then take the
$\Lambda_W \to 0$ limit. This does not change the UV properties of
the theory, but it alters the IR properties. The deformed action of
the non-relativistic renormalizable gravitational theory  is given
by\cite{KS}
\begin{eqnarray}
S_{HL}&=&\int dtd^3x, \Big({\cal L}_0 + \tilde{{\cal L}}_1\Big)\,\\
{\cal L}_0 &=& \sqrt{g}N\left\{\frac{2}{\kappa^2}(K_{ij}K^{ij}
\label{action1}-\lambda K^2)+\frac{\kappa^2\mu^2(\Lambda_W R
  -3\Lambda_W^2)}{8(1-3\lambda)}\right\}\,,\\ \tilde{{\cal L}}_1&=&
\sqrt{g}N\left\{\frac{\kappa^2\mu^2 (1-4\lambda)}{32(1-3\lambda)}R^2
-\frac{\kappa^2}{2w^4} \left(C_{ij} -\frac{\mu w^2}{2}R_{ij}\right)
\left(C^{ij} -\frac{\mu w^2}{2}R^{ij}\right) +\mu^4R
\right\}.\label{action2}
\end{eqnarray}
Here $C_{ij}$ is the Cotton tensor, defined by
\begin{eqnarray}
C^{ij}=\epsilon^{ik\ell}\nabla_k\left(R^j{}_\ell
-\frac14R\delta_\ell^j\right)=\epsilon^{ik\ell}\nabla_k R^j{}_\ell
-\frac14\epsilon^{ikj}\partial_kR\,.\label{def.K.C}
\end{eqnarray}
Comparing the action to that of general relativity in the ADM
formalism, one can find that the speed of light, Newton's constant
and the cosmological constant are given by
\begin{eqnarray}
 c=\frac{\kappa^2\mu}{4}
\sqrt{\frac{\Lambda_W}{1-3\lambda}},\;\;\;\;
G=\frac{\kappa^2}{32\pi\,c}\,,\qquad \Lambda=\frac{3}{2}
\Lambda_W\,.\label{cg}
\end{eqnarray}
Taking $N^i=0$, the spherically symmetric solutions could  be
obtained with the metric
ansatz~\cite{LMP,CCO,CLS,CY,MK,Nis,CCO1,Gho}
\begin{eqnarray}
\label{ssm} ds^2 = - N^2(r)\,dt^2 + \frac{dr^2}{f(r)} + r^2
(d\theta^2 +\sin^2\theta d\phi^2)\,
\end{eqnarray}
Substituting the metric ansatz (\ref{ssm}) into the action, and then
vary the functions $N$ and $f$, one can find that the reduced
Lagrangian reads
\begin{eqnarray}
\label{react} \tilde{{\cal L}}_1=\frac{ \kappa^2\mu^2 N
}{8(1-3\lambda)\sqrt{f}}\Bigg(  \frac{\lambda-1}{2} f'^2 -
\frac{2\lambda (f-1)}{ r}f' + \frac{(2\lambda-1)(f-1)^2}{ r^2}-2w(1
- f - r f')\Bigg)
\end{eqnarray} with
$w=8\mu^2(3\lambda-1)/\kappa^2$. For
$\lambda=1~(w=16\mu^2/\kappa^2)$, we have a solution where $f$ and
$N$ are determined to be
\begin{eqnarray}
\label{sol1}
N^2=f=\frac{2(r^2-2Mr+\alpha)}{r^2+2\alpha+\sqrt{r^4+8\alpha Mr}},
\end{eqnarray}
where $\alpha=1/(2w)$ and $M$ is an integration constant related to
the mass. The metric of this black hole looks like that of
Gauss-Bonnet black hole. The event horizons are given by
\begin{eqnarray}
r_\pm=M\pm \sqrt{M^2-\alpha},
\end{eqnarray}
and the Hawking temperature is
\begin{eqnarray}
T_H=\frac{f'}{4\pi}\mid_{r=r_+}= \frac{\sqrt{M^2-\alpha}}{\pi(
r_+^2+2\alpha+\sqrt{r^4_++8\alpha Mr_+})}.
\end{eqnarray}
The thermodynamics of this black hole has been studied in
\cite{Myung,myung1,sj}. Myung \cite{myung1} has calculated the ADM
mass of this black hole and find that it is similar to that of
$4$-dimensional Einstein-Born-Infeld black hole \cite{bobh}. It is
well known that the Born-Infeld electrodynamics is one of the
important nonlinear electromagnetic theories. As the Born-Infeld
scale parameter $b$ tends to zero, the Einstein-Maxwell theory is
recovered and the Einstein-Born-Infeld black hole is reduced to
Reissner-Norstr\"{om} black hole. Applying such kind of ADM mass
proposed in \cite{myung1}, Wang \textit{et al} \cite{sj} find that
the integral and differential forms of the first law of
thermodynamics are still valid for the deformed Ho\v{r}ava-Lifshitz
black hole. The potentially observable properties of this black hole
were considered in \cite{s1,rak,sb1}. In this letter, our main
purpose is to study the quasinormal modes of massless scalar field
in this spacetimes and to see what there exists some new feature in
the dynamical evolution of the perturbation in the black hole in
Ho\v{r}ava-Lifshitz theory.

The Klein-Gordon equation for a massless scalar field in this
spacetime is
\begin{eqnarray}
\frac{1}{\sqrt{-g}}\partial_{\mu}(\sqrt{-g}g^{\mu\nu}\partial_{\nu})
\psi=0.\label{WE}
\end{eqnarray}
Separating $\psi=e^{-i\omega t}R(r)Y_{lm}(\theta,\phi)/r$, we can
obtain the radial equation for the scalar perturbation in the
deformed Ho\v{r}ava-Lifshitz black hole spacetime
\begin{eqnarray}
\frac{d^2 R(r)}{dr^2_*}+[\omega^2-V(r)]R(r)=0,\label{jw}
\end{eqnarray}
where $r_*$ is the tortoise coordinate (which is defined by
$dr_*=\frac{1}{f}dr$) and the effective potential $V(r)$ reads
\begin{eqnarray}
V(r)=\frac{2(r^2-2Mr+\alpha)}{r^2+2\alpha+\sqrt{r^4+8\alpha
Mr}}\bigg[\frac{l(l+1)}{r^2}+\frac{1}{\alpha}\bigg(1-\frac{r^3+2\alpha
M}{r\sqrt{r^4+8\alpha Mr}}\bigg)\bigg].\label{efp}
\end{eqnarray}
Obviously, as $\alpha=0$ the effective potential $V(r)$ can be
reduced to that of the Schwarzschild black hole. When $\alpha$
increases we find that the peak value of the potential barrier gets
lower for $l=0$  and higher for $l=1$, which is shown in figure (1).
In the Reissner-Norstr\"{om} black hole background, for all $l$ the
peak value of the effective potential increases with the charge $q$
of the black hole. This means that although the formulas of the
outer and inner horizons are very similar, the behaviors of the
effective potential are different in these black hole background. In
the Einstein-Born-Infeld black hole, the variety of the effective
potential with the charge $q$ is similar to that in
Reissner-Norstr\"{om} black hole spacetime. However, the Born-Infeld
scale parameter $b$ decreases the peak value of $V(r)$ for all $l$.
These results imply the quasinormal modes in the deformed
Ho\v{r}ava-Lifshitz black hole possess some different properties
from those of the black holes in the Einstein-Maxewell and
Einstein-Born-Infeld gravities.
\begin{figure}[ht]
\begin{center}
\includegraphics[width=5cm]{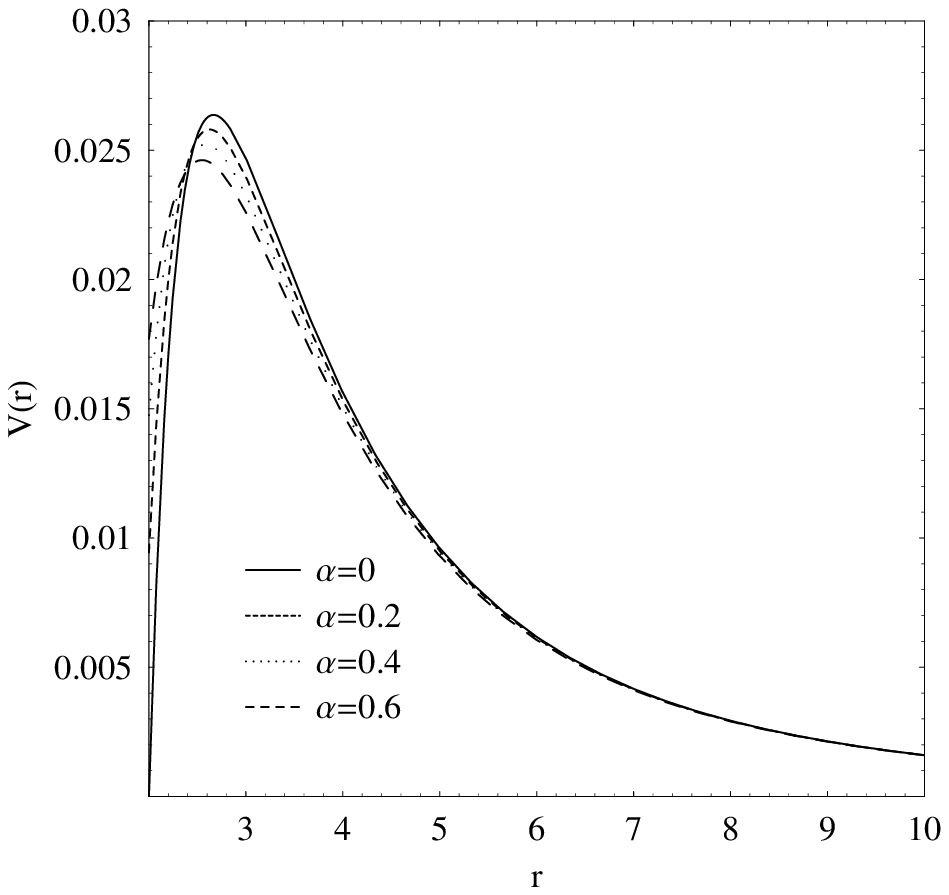}\includegraphics[width=5cm]{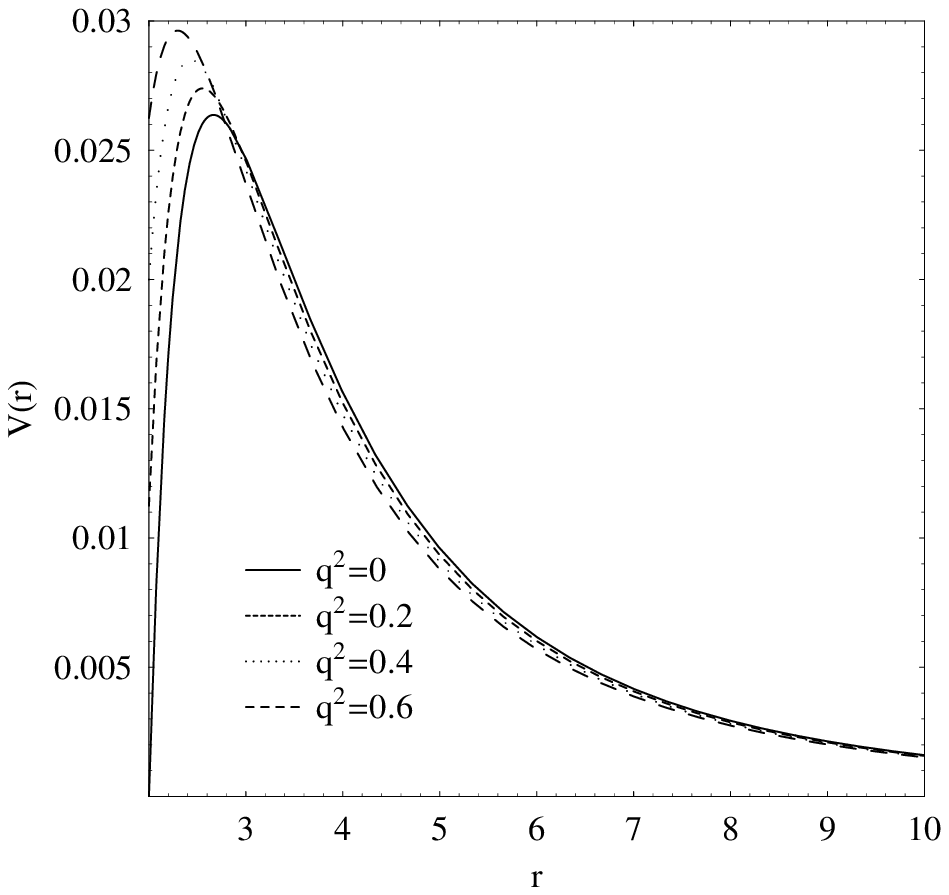}\includegraphics[width=5cm]{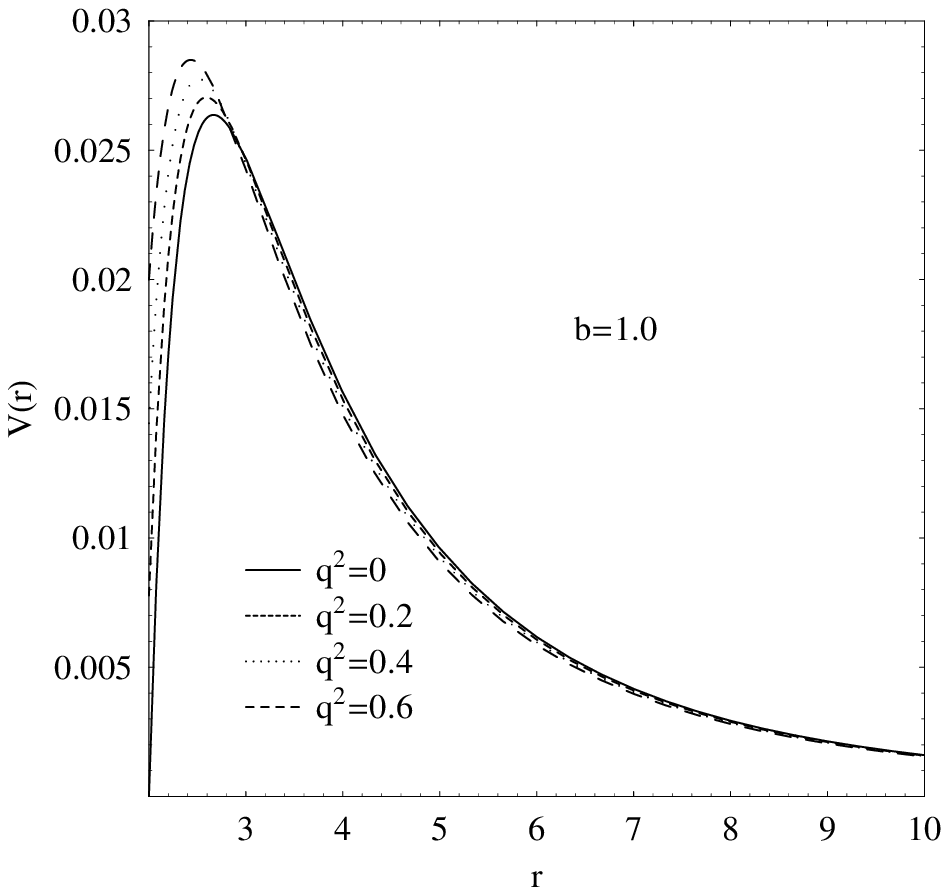}\\
\includegraphics[width=5cm]{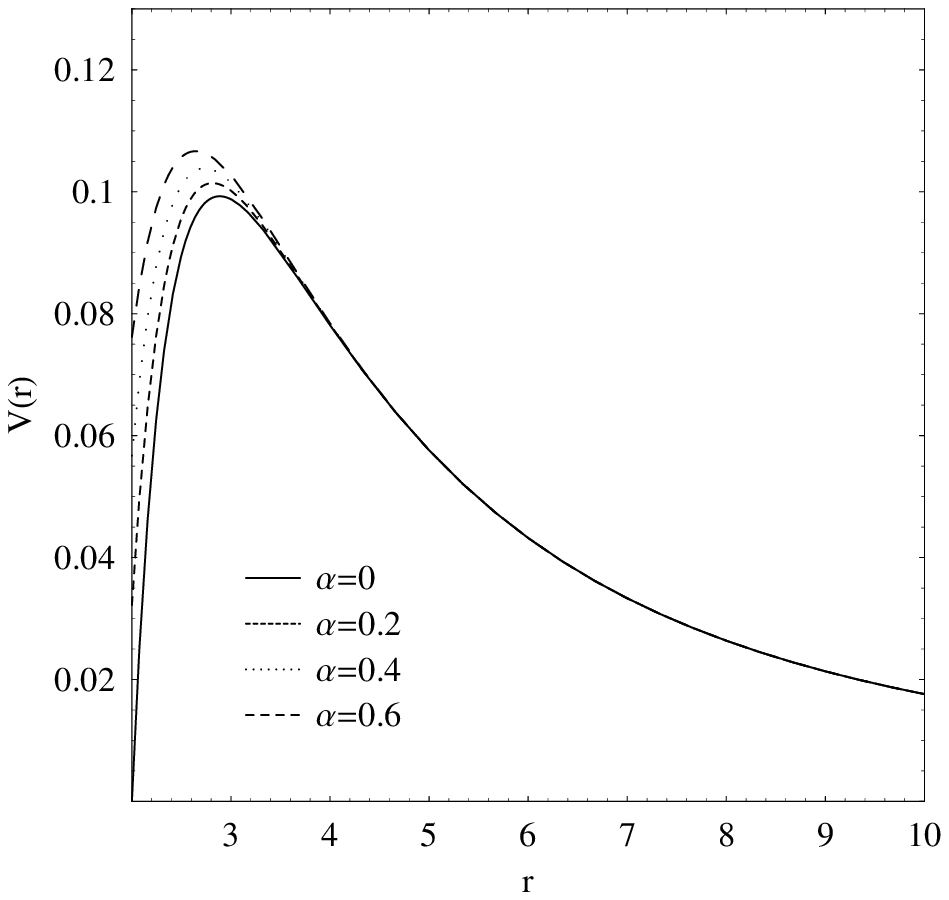}\includegraphics[width=5cm]{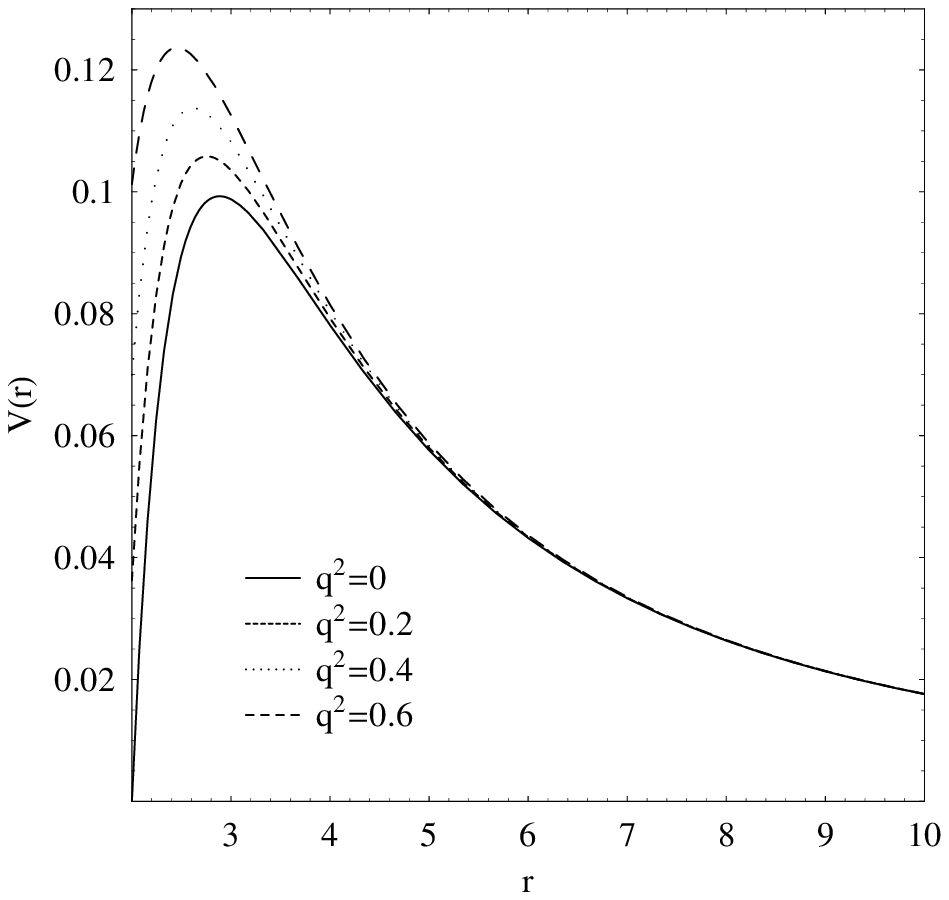}\includegraphics[width=5cm]{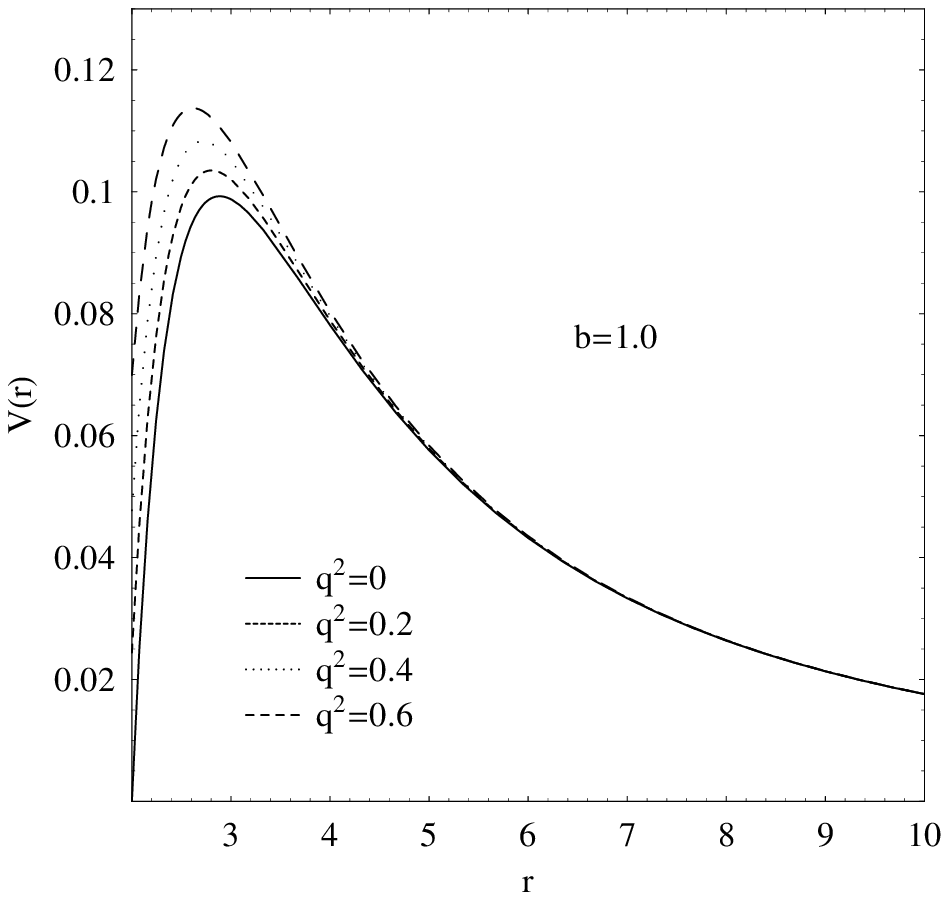}
\caption{Variety of the effective potential $V(r)$ with the polar
coordinate $r$ in the deformed Ho\v{r}ava-Lifshitz black hole (left
column), Reissner-Norstr\"{om} black hole (middle column) and
Einstein-Born-Infeld black hole with fixed scale parameter $b=1.0$
(right column). The figures in the upper row are for $l=0$ and in
the lower row are for $l=1$ and $M=1$.}
\end{center}
\label{fig1}
\end{figure}

We are now in a position to apply the the third-order WKB
approximation method approximation to evaluate the fundamental
quasinormal modes ($n=0$) of massless scalar perturbation in the
deformed Ho\v{r}ava-Lifshitz black hole. We expect to see what
effects of Ho\v{r}ava-Lifshitz parameter $\alpha$ can be reflected
in the quasinormal modes' behavior. The formula for the complex
quasinormal frequencies $\omega$ in this approximation is given by
\cite{wkb,wkba1,wkba2}
\begin{eqnarray}
\omega^2=[V_0+(-2V^{''}_0)^{1/2}\Lambda]-i(n+\frac{1}{2})(-2V^{''}_0)^{1/2}(1+\Omega),
\end{eqnarray}
where
\begin{eqnarray}
\Lambda&=&\frac{1}{(-2V^{''}_0)^{1/2}}\left\{\frac{1}{8}\left(\frac{V^{(4)}_0}{V^{''}_0}\right)
\left(\frac{1}{4}+\alpha^2\right)-\frac{1}{288}\left(\frac{V^{'''}_0}{V^{''}_0}\right)^2
(7+60\alpha^2)\right\},\nonumber\\
\Omega&=&\frac{1}{(-2V^{''}_0)}\bigg\{\frac{5}{6912}
\left(\frac{V^{'''}_0}{V^{''}_0}\right)^4
(77+188\alpha^2)\nonumber\\&-&
\frac{1}{384}\left(\frac{V^{'''^2}_0V^{(4)}_0}{V^{''^3}_0}\right)
(51+100\alpha^2)
+\frac{1}{2304}\left(\frac{V^{(4)}_0}{V^{''}_0}\right)^2(67+68\alpha^2)
\nonumber\\&+&\frac{1}{288}
\left(\frac{V^{'''}_0V^{(5)}_0}{V^{''^2}_0}\right)(19+28\alpha^2)-\frac{1}{288}
\left(\frac{V^{(6)}_0}{V^{''}_0}\right)(5+4\alpha^2)\bigg\},
\end{eqnarray}
and
\begin{eqnarray}
\alpha=n+\frac{1}{2},\;\;\;\;\;
V^{(s)}_0=\frac{d^sV}{dr^s_*}\bigg|_{\;r_*=r_*(r_{p})} \nonumber,
\end{eqnarray}
$n$ is overtone number and $r_{p}$ is the value of polar coordinate
$r$ corresponding to the peak of the effective potential
(\ref{efp}). Setting $M=1$ and substituting the effective potential
(\ref{efp}) into the formula above, we can obtain the quasinormal
frequencies of scalar perturbation in the deformed
Ho\v{r}ava-Lifshitz black hole.
\begin{table}[h]
\begin{center}
\begin{tabular}[b]{cccc}
 \hline \hline
 \;\;\;\; $\alpha$ \;\;\;\; & \;\;\;\; $\omega\ \ \ (l=0)$\;\;\;\;  & \;\;\;\;  $\omega \ \ \ (l=1)$\;\;\;\;
 & \;\;\;\; $\omega \ \ \ (l=2)$ \;\;\;\; \\ \hline
\\
0.0& \;\;\;\;\;0.104647-0.115197i\;\;\;\;\;  & \;\;\;\;
0.291114-0.098001i\;\;\;\;\;
 & \;\;\;\;\;0.483211-0.096805i\;\;\;\;\;
 \\
0.1&0.104687-0.111855i&0.293313-0.096159i&0.486927-0.095211i
 \\
0.2&0.104300-0.108459i&0.295640-0.094238i&0.490850-0.093510i
 \\
0.3&0.103481-0.105037i&0.298101-0.092219i&0.495005-0.091681i
\\
0.4&0.102257-0.101646i&0.300704-0.090070i&0.499417-0.089697i
\\
0.5&0.100690-0.098374i&0.303454-0.087770i&0.504120-0.087522i
\\
\hline \hline
\end{tabular}
\end{center}
\caption{The fundamental ($n=0$) quasinormal frequencies of scalar
field in the in the deformed Ho\v{r}ava-Lifshitz black hole for
$l=0$, $1$, $2$.}
\end{table}
\begin{table}[h]
\begin{center}
\begin{tabular}[b]{cccc}
 \hline \hline
 \;\;\;\; $q$ \;\;\;\; & \;\;\;\; $\omega\ \ \ (l=0)$\;\;\;\;  & \;\;\;\;  $\omega \ \ \ (l=1)$\;\;\;\;
 & \;\;\;\; $\omega \ \ \ (l=2)$ \;\;\;\; \\ \hline
\\
0.0& \;\;\;\;\;0.104647-0.115197i\;\;\;\;\;  & \;\;\;\;
0.291114-0.098001i\;\;\;\;\;
 & \;\;\;\;\;0.483211-0.096805i\;\;\;\;\;
 \\
0.1&0.104856-0.115204i&0.291616-0.098051i&0.484025-0.096857i
 \\
0.2&0.105493-0.115221i&0.293145-0.098196i&0.486503-0.097011i
 \\
0.3&0.106586-0.115230i&0.295770-0.098430i&0.490758-0.097261i
\\
0.4&0.108179-0.115195i&0.299617-0.098737i&0.496997-0.097593i
\\
0.5&0.110339-0.115049i&0.304891-0.099084i&0.505561-0.097977i
\\
\hline \hline
\end{tabular}
\end{center}
\caption{The fundamental ($n=0$) quasinormal frequencies of scalar
field in the Reissner-Norstr\"{om} black hole for $l=0$, $1$, $2$.}
\end{table}

\begin{table}[h]
\begin{center}
\begin{tabular}[b]{cccc}
 \hline \hline
 \;\;\;\; $q$ \;\;\;\; & \;\;\;\; $\omega\ \ \ (l=0)$\;\;\;\;  & \;\;\;\;  $\omega \ \ \ (l=1)$\;\;\;\;
 & \;\;\;\; $\omega \ \ \ (l=2)$ \;\;\;\; \\ \hline
\\
0.0& \;\;\;\;\;0.104647-0.115197i\;\;\;\;\;  & \;\;\;\;
0.291114-0.0980010i\;\;\;\;\;
 & \;\;\;\;\;0.483211-0.0968050i\;\;\;\;\;
 \\
0.1&0.104786-0.115202i&0.291449-0.0980342i&0.483753-0.0968396i
 \\
0.2&0.105209-0.115215i&0.292462-0.0981318i&0.485395-0.0969430i
 \\
0.3&0.105927-0.115228i&0.294184-0.0982911i&0.488185-0.0971125i
\\
0.4&0.106961-0.115229i&0.296667-0.0985064i&0.492212-0.0973430i
\\
0.5&0.108344-0.115194i&0.299995-0.0987669i&0.497609-0.0976251i
\\
\hline \hline
\end{tabular}
\end{center}
\caption{The fundamental ($n=0$) quasinormal frequencies of scalar
field in the Einstein-Born-Infeld black hole with $b=1.0$ for $l=0$,
$1$, $2$.}
\end{table}

\begin{table}[h]
\begin{center}
\begin{tabular}[b]{cccc}
 \hline \hline
 \;\;\;\; $b$ \;\;\;\; & \;\;\;\; $\omega\ \ \ (l=0)$\;\;\;\;  & \;\;\;\;  $\omega \ \ \ (l=1)$\;\;\;\;
 & \;\;\;\; $\omega \ \ \ (l=2)$ \;\;\;\; \\ \hline
\\
0.0& \;\;\;\;\;0.105493-0.115221i\;\;\;\;\;  & \;\;\;\;
0.293145-0.0981960i\;\;\;\;\;
 & \;\;\;\;\;0.486503-0.0970110i\;\;\;\;\;
 \\
0.1&0.105465-0.115221i&0.293077-0.0981895i&0.486392-0.0970044i
 \\
0.2&0.105436-0.115220i&0.293008-0.0981831i&0.486280-0.0969976i
 \\
0.3&0.105408-0.115220i&0.292940-0.0981768i&0.486169-0.0969908i
\\
0.4&0.105379-0.115219i&0.292871-0.0981704i&0.486058-0.0969840i
\\
0.5&0.105351-0.115218i&0.292803-0.0981640i&0.485947-0.0969772i
\\
\hline \hline
\end{tabular}
\end{center}
\caption{The fundamental ($n=0$) quasinormal frequencies of scalar
field in the Einstein-Born-Infeld black hole with $q=0.2$ for $l=0$,
$1$, $2$.}
\end{table}

\begin{figure}[ht]
\begin{center}
\includegraphics[width=6cm]{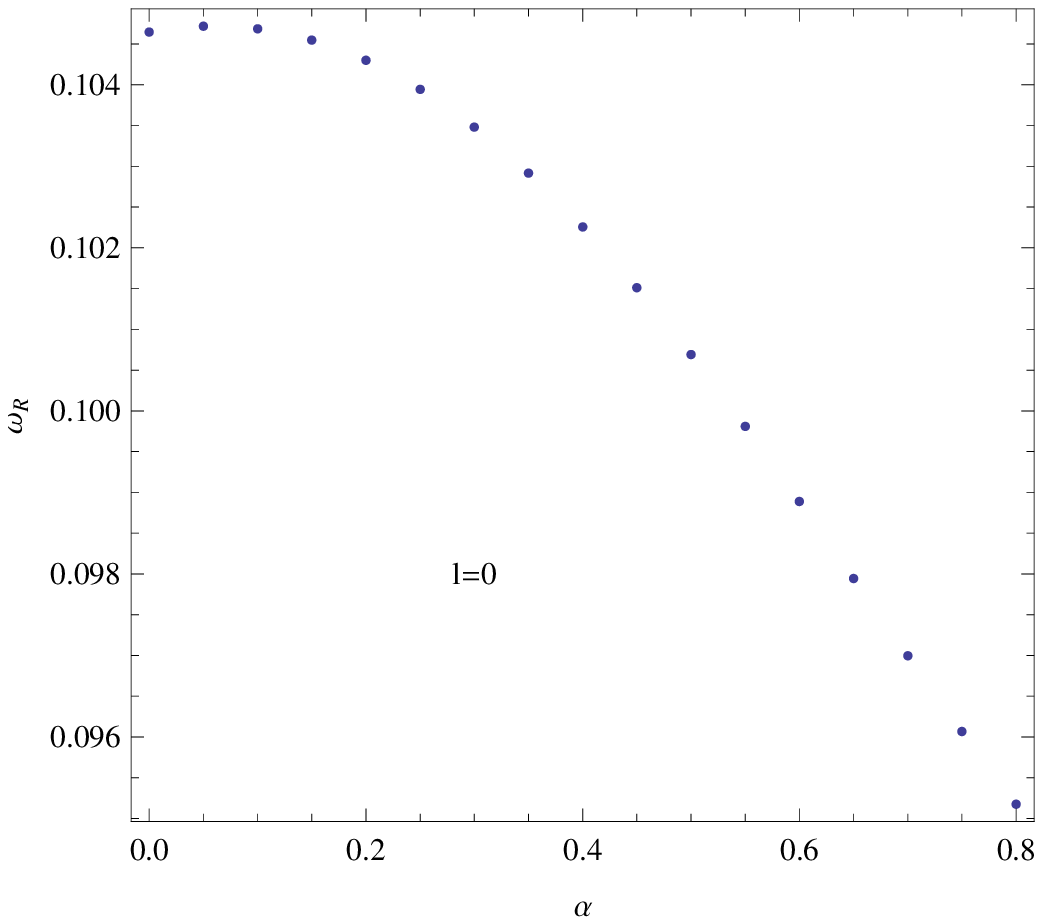}\;\;\;\includegraphics[width=6cm]{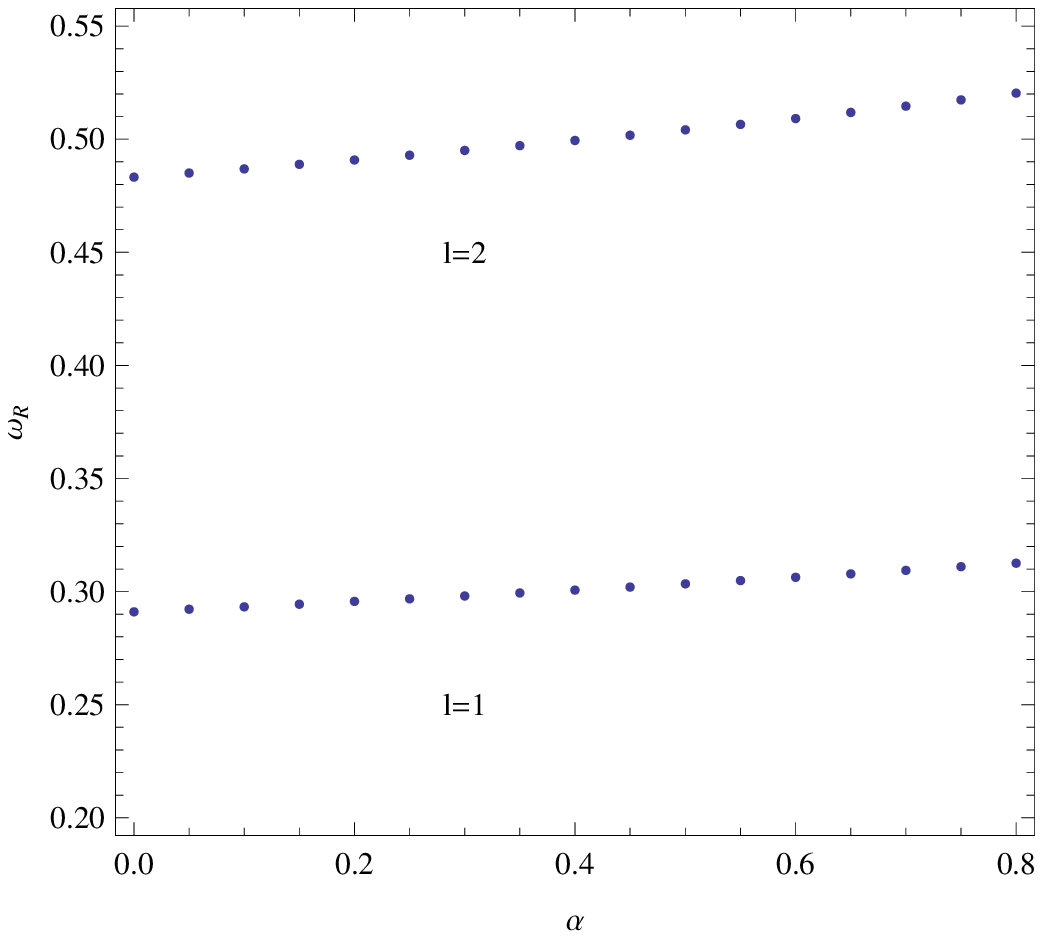}
\caption{Variety of the real part of the fundamental quasinormal
modes with $\alpha$ in the deformed Ho\v{r}ava-Lifshitz black hole
spacetime. The left is for $l=0$ and the right for $l=1,\;2$.}
\end{center}
\label{fig2}
\end{figure}
\begin{figure}[ht]
\begin{center}
\includegraphics[width=6cm]{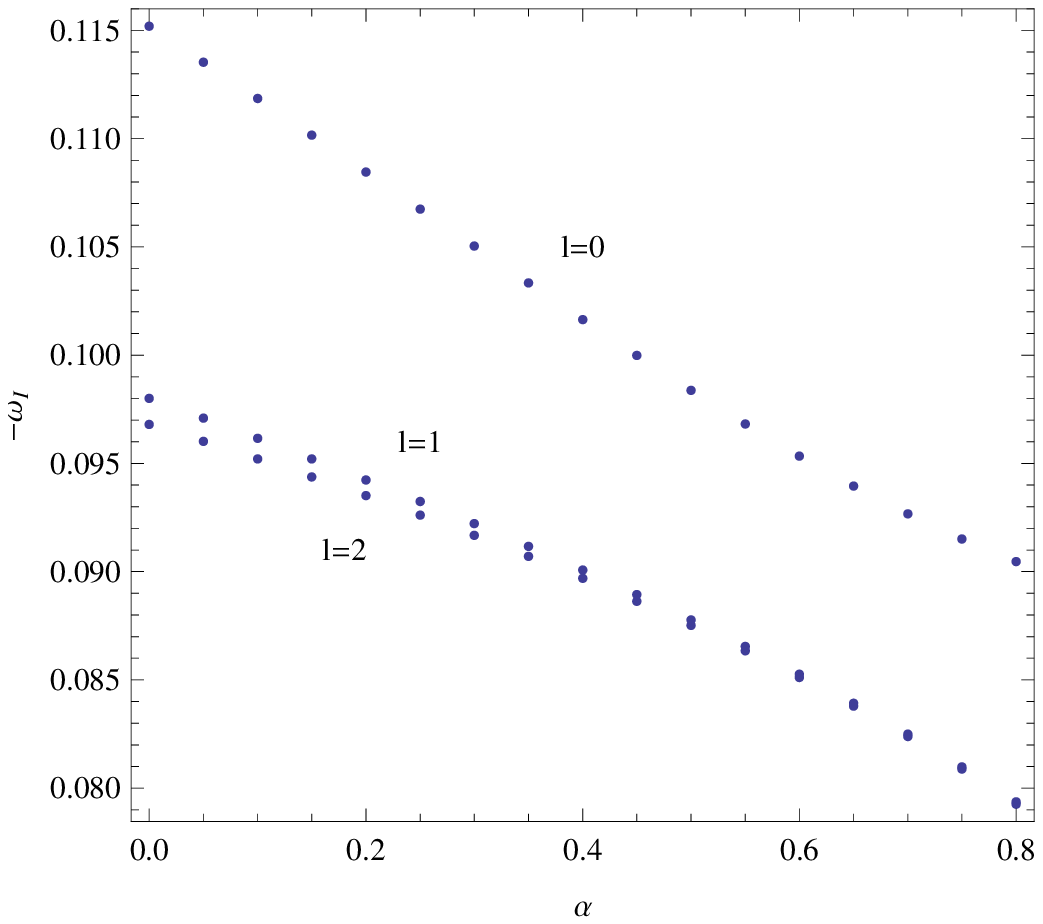}
\caption{Variety of the absolute value of imaginary parts of the
fundamental quasinormal modes with $\alpha$ for fixed $l=1,\;2$in
the deformed Ho\v{r}ava-Lifshitz black hole spacetime.}
\end{center}
\label{fig3}
\end{figure}

In tables (I)-(IV), we list the fundamental quasinormal frequencies
of the massless scalar perturbation field for fixed $l=0$, $1$ and
$2$ in the deformed Ho\v{r}ava-Lifshitz, the Reissner-Norstr\"{om}
and Einstein-Born-Infeld black hole spacetimes, respectively. From
the table (I) and figures (2) and (3), we find that with the
increase of the parameter $\alpha$ the real parts decrease for $l=0$
and increase for $l=1 $ and $l=2$. The absolute value of imaginary
parts for all $l$ decrease. Comparing with the quasinormal modes of
the Reissner-Norstr\"{om} and Einstein-Born-Infeld black holes
\cite{RN1,RN2,RN3,qbI}, we find from the tables (I)-(III) that the
dependence of quasinormal modes on the parameter $\alpha$ is
different from that on the charge $q$ in the usual
Reissner-Norstr\"{om} and Einstein-Born-Infeld black hole
spacetimes. With the increase of $q$, the real part of the
quasinormal modes increases in both Reissner-Norstr\"{om} and
Einstein-Born-Infeld black holes. While the absolute value of
imaginary parts first increase and then decrease. Moreover, from
tables (I) and (IV), we also find the effect of the parameter
$\alpha$ on quasinormal modes is different from that of the
Born-Infeld parameter $b$ in the Einstein-Born-Infeld black hole. As
$b$ increases, both the real parts and the absolute value of
imaginary parts decrease for all $l$. Moreover, the speed of
decrease of the absolute value of imaginary parts with $b$ in the
Einstein-Born-Infeld black hole is smaller than that with $\alpha$
in the deformed Ho\v{r}ava-Lifshitz black hole. The tables (I)-(IV)
also tell us that the absolute value of imaginary parts is smaller
than that of in the Schwarzschild, Reissner-Norstr\"{om} and
Einstein-Born-Infeld black hole spacetimes. This means that the
scalar perturbation decays more slowly in the deformed
Ho\v{r}ava-Lifshitz black hole.

For the late-time evolution of the massless scalar perturbation, one
can expect that it has the same form as that in background of a
Schwarzschild black hole because that in the deformed
Ho\v{r}ava-Lifshitz black hole spacetimes the effective potentials
have the same asymptotic behaviors when the polar coordinate $r$
approaches to infinite.

In summary we study the quasinormal modes of the massless scalar
perturbation in the background of a deformed black hole in the
Ho\v{r}ava-Lifshitz gravity. Our results show that the dependence of
quasinormal modes on the parameter $\alpha$ is different from that
on the charge $q$ in the usual Reissner-Norstr\"{om} black hole
spacetime. Our analysis also indicate that the effect of the
parameter $\alpha$ on quasinormal modes is different from that of
the Born-Infeld parameter $b$ in the Einstein-Born-Infeld black
hole. Moreover, we also find the absolute value of imaginary parts
is smaller than that of in the Schwarzschild, Reissner-Norstr\"{om}
and Einstein-Born-Infeld black hole spacetimes, which means that the
scalar perturbation decays more slowly in the deformed
Ho\v{r}ava-Lifshitz black hole. This would open a window to examine
the Ho\v{r}ava-Lifshitz theory, Einstein-Maxewell theory and
Einstein-Born-Infeld theory in the future. It would be of interest
to generalize our study to the other Ho\v{r}ava-Lifshitz black hole
spacetimes in which $\lambda\neq1$, such as the black holes in refs
\cite{CY,LMP,CCO,CLS,Gho}. Work in this direction will be reported
in the future.

\begin{acknowledgments}

This work was partially supported by the National Natural Science
Foundation of China under Grant No.10875041; the Scientific Research
Fund of Hunan Provincial Education Department Grant No.07B043 and
the construct program of key disciplines in Hunan Province. J. L.
Jing's work was partially supported by the National Natural Science
Foundation of China under Grant No.10675045; the FANEDD under Grant
No. 200317; the National Basic Research of China under Grant No.
2010CB833004; and the Hunan Provincial Natural Science Foundation of
China under Grant No.08JJ3010.
\end{acknowledgments}

\end{document}